\documentclass[aps,pre,twocolumn,groupedaddress,showpacs]{revtex4}
\usepackage{graphicx}
\usepackage{amssymb}

\begin{document}
%\preprint{}
\title{Coarsening of Two Dimensional Foam on a Dome}
\author{A. E. Roth, C. D. Jones, and D. J. Durian}
\affiliation{Department of Physics \& Astronomy, University of Pennsylvania, Philadelphia, PA 19104-6396, USA}

\date{\today}

\begin{abstract}
In this paper we report on bubble growth rates and on the statistics of bubble topology for the coarsening of a dry foam contained in the narrow gap between two hemispheres.  By contrast with coarsening in flat space, where six-sided bubbles neither grow nor shrink, we observe that six-sided bubbles grow with time at a rate that depends on their size.  This result agrees with the modification to von~Neumann's law predicted by J.~E.~Avron and D.~Levine.  For bubbles with a different number of sides, except possibly seven, there is too much noise in the growth rate data to demonstrate a difference with coarsening in flat space.  In terms of the statistics of bubble topology, we find fewer 3, 4, and 5 sided bubbles, and more 6 and greater sided bubbles, in comparison with the stationary distribution for coarsening in flat space.  We also find good general agreement with the Aboav-Weaire law for the average number of sides of the neighbors of an $n$-sided bubble.
\end{abstract}

\pacs{82.70.Rr}

\maketitle

%=========================================================================================

\section{Introduction}

Coarsening is a process in foams by which there is diffusion of gas across films such that some bubbles grow and other bubbles shrink.  This progresses in such a way that the average bubble area increases over time \cite{WeaireHutzlerBook}.  Coarsening is not limited to foams, and is also relevant in other systems involving domain growth \cite{GlazierWeaire92, Stavans93}.  For an ideal dry two dimensional foam, John von~Neumann showed that the rate of change of area of a bubble in a two dimensional foam is \cite{VonNeumann}:
\begin{equation}
\frac{\textrm{d}A}{\textrm{d}t} = K_o (n-6),
\label{VN}
\end{equation}
where $n$ is the number of sides of a bubble, and $K_o$ is a constant of proportionality.  Remarkably, the shape of the bubble, its edge lengths, and its set of neighbors, all do not matter.

In 1992 Avron and Levine \cite{AvronLevine92} generalized von~Neumann's law to predict the rate of area change for bubbles coarsening on a curved surface.  The essential ingredient is that the sum of turning angles around each bubble is no longer $2 \pi$, as in flat space, but rather depends on the integral of Gaussian curvature, $\kappa_{G}$, over the bubble area.  This modifies the von~Neumann law to:
\begin{equation}
\frac{\textrm{d}A}{{\rm d}t} = K_o \left[(n-6) + \frac{3}{\pi} \int \kappa_{G} {\rm d}A\right].
\label{AvronLevine}
\end{equation}
In the case of a surface of constant positive curvature, such as a dome of radius $R$, this reduces to
\begin{equation}
\frac{\textrm{d}A}{\textrm{d}t} = K_o \left[(n-6) + \frac{3 A}{\pi R^{2}}\right]
\label{AvronLevine2}
\end{equation}
The rate of change of bubble area thus depends on the number of sides {\it and the area} of the bubble.

There have been numerous theoretical and simulation studies of coarsening for foams in two dimensional flat space \cite{Beenakker86, KermodeWeaire89, GlazierAndersonGrest90, HerdtleAref92, Segeletal93, Flyvbjerg93, Stavans93sf, NeuSch97, Rutenberg05}.  But to date we are aware of only one simulation that includes the effect of substrate curvature \cite{PeczakGrestLevine93}.  There, the authors used a modified Potts model for two dimensional foam coarsening on spheres, toroids, and pseudospheres.  For spheres, they focused on how the area distribution and average area change over time, and find that at late times the dynamics are dominated by the appearance of `singular bubbles' much larger than the average that quickly grow to cover the sphere.  There is minimal discussion of the coarsening of individual bubbles, and no discussion of side number or other distributions of the system.

While the coarsening of foams in two dimensional flat space has been well-measured \cite{GlazierGrossStavans87, StavansGlazier89, Stavans90, KnoblerPRA90, Bergeetal90, StavansKrichevsky92, Stavans93sf, Icaza94, RosaFortes99, RosaFortes02, RothDurian11}, we are unaware of any experiments to test the modified law of Avron and Levine for foam in two dimensional curved space.  However metallurgical grain growth on curved substrates has been reported.  In Ref.~\cite{Brockman}, the results are said to be preliminary and no growth rate data are shown.  In Ref.~\cite{Glicksman}, the deviation from the coarsening rate for flat space is masked by noise, but statistical analysis is reported to demonstrate consistency with Eq.~(\ref{AvronLevine}).  In this paper we use a hemispheric  cell to create a curved two dimensional foam.  We use image analysis to track individual bubbles and measure bubble dynamics such as coarsening rate.  We also measure bubble statistics, such as the distribution of number of sides and compare this to results from a flat cell.  Our image quality and analysis methods are sufficient to demonstrate directly, for six-sided bubbles, that the growth rates are different from flat space and are consistent with Avron and Levine \cite{AvronLevine92}.

\section{Materials and Methods}

To measure coarsening rates of two dimensional foams on a curved surface, we constructed a cell from two hemispherical polycarbonate domes.  The smaller dome has an outer diameter of 12.5~cm, and the larger dome has inner diameter 13.3~cm, creating a 4~mm gap.  The smaller dome was glued to a flat acrylic plate.  The larger dome was placed over the smaller dome and separated from the plate by an O-ring of diameter 0.25~inches.  We were careful to ensure that the two domes were aligned concentrically.  The upper dome was then screwed to the plate to create a sealed chamber of constant curvature.

The solution we used to create our foam was a liquid consisting of 75\% deionized water, 20\% glycerin, and 5\% Dawn Ultra Concentrated dishwashing liquid.  This created a foam that was stable and generally lasted many days.  The foam was prepared by putting 35~mL of solution into the chamber (this fills the dome to about 2~cm above the O-ring) and shaking it until a uniform opaque foam was created, with an average bubble size much less than the separation of the domes.  The chamber was then left to coarsen until a single layer of bubbles remained between the two domes.  This took about 24~hours.  Two dimensional coarsening could then be observed for the next 2-4 days.  Film ruptures were sometimes observed at the end of this period.  Bubble statistics were not taken after any ruptures, though single bubble dynamics were still considered.

\begin{figure}
\includegraphics[width=3.00in]{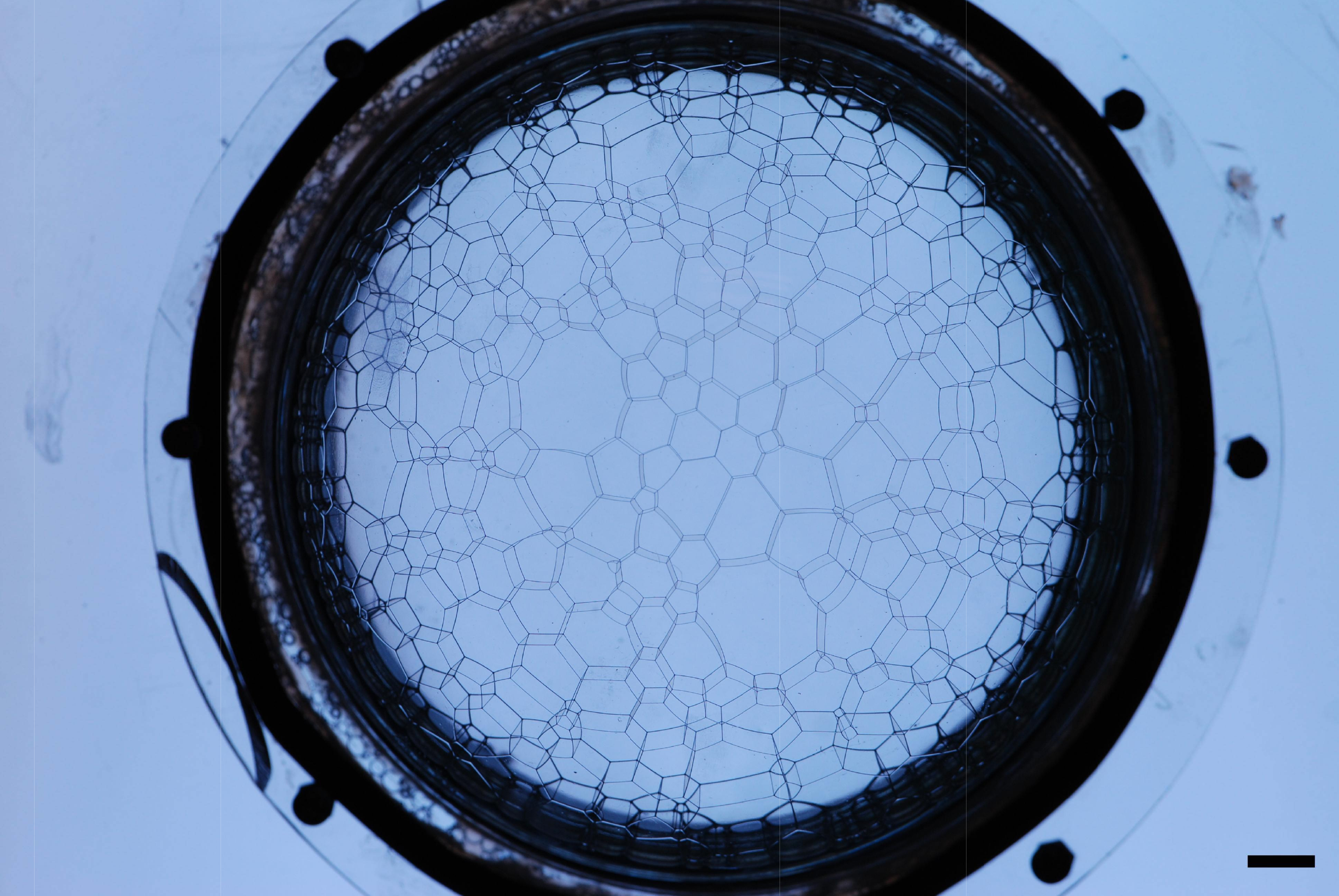}
\caption{Sample photograph of a two-dimensional foam, coarsening between nested polycarbonate hemispheres with a 4~mm gap.  The scale bar is 1~cm.}
\label{dome}
\end{figure}

To photograph the foam, the chamber was placed 65~cm above a Vista Point A lightbox, which provided a spatially and temporally uniform light source.  A Nikon D80 camera with a Nikkor AF-S 300 mm 1:2.8 D lens was mounted 2.5~m above the chamber.  The camera was controlled by a computer to take pictures every two minutes.  The apparatus was left to collect pictures, for a period ranging from a few days up to a week.  This process was repeated three times to build up statistics.  A sample photograph can be seen in Fig.~\ref{dome}.  Note in this photograph that, especially towards the edge, it is possible to see both the Plateau borders on the top dome, as well as the Plateau borders on the bottom dome.  This makes it difficult to identify the correct boundaries of the cells.  We address this issue as part of our image analysis.

After we have taken a series of images of the dome, we perform analysis to get out the true areas of the bubbles on the dome.  Note that the images constitute an orthographic projection of a sphere (or hemisphere) onto a plane, where the point of projection is infinity.  This projection converts the positions according to the following equations \cite{MapBook}:
\begin{equation}
\begin{array}{l}
x = R \cos\varphi \sin\lambda\\
y = R \sin\varphi
\end{array}
\label{orthographic}
\end{equation}
where $R$ is the radius of the sphere, $\lambda$ is the longitude, $\varphi$ is the latitude, and the center of the domes is defined as $\lambda = 0, \varphi = 0$.  The first problem is that in a given image, both the Plateau borders on the top and bottom domes are visible.  In order to isolate a single set of Plateau borders so that the cells' edges are defined correctly, we recognize that the image is projected in two ways.  The Plateau borders on the top dome are an orthographic projection of the foam using the radius of the top dome, and the Plateau borders on the bottom dome are an orthographic projection of the foam using the radius of the bottom dome; both are combined into the same image.  To undo this transformation, we take the inverse of the transformation twice, once using the radius for the top dome and once using the radius for the bottom dome.  The resulting two images are thresholded and dilated.  The images are then multiplied.  This kills the Plateau borders that do not correspond to the transformation.  That is, the Plateau borders from the bottom dome that were transformed using the radius of the top dome are killed and vice versa.  The result is a binary image with the correct latitudes and longitudes of the Plateau borders on the dome.

\begin{figure}
\includegraphics[width=3.00in]{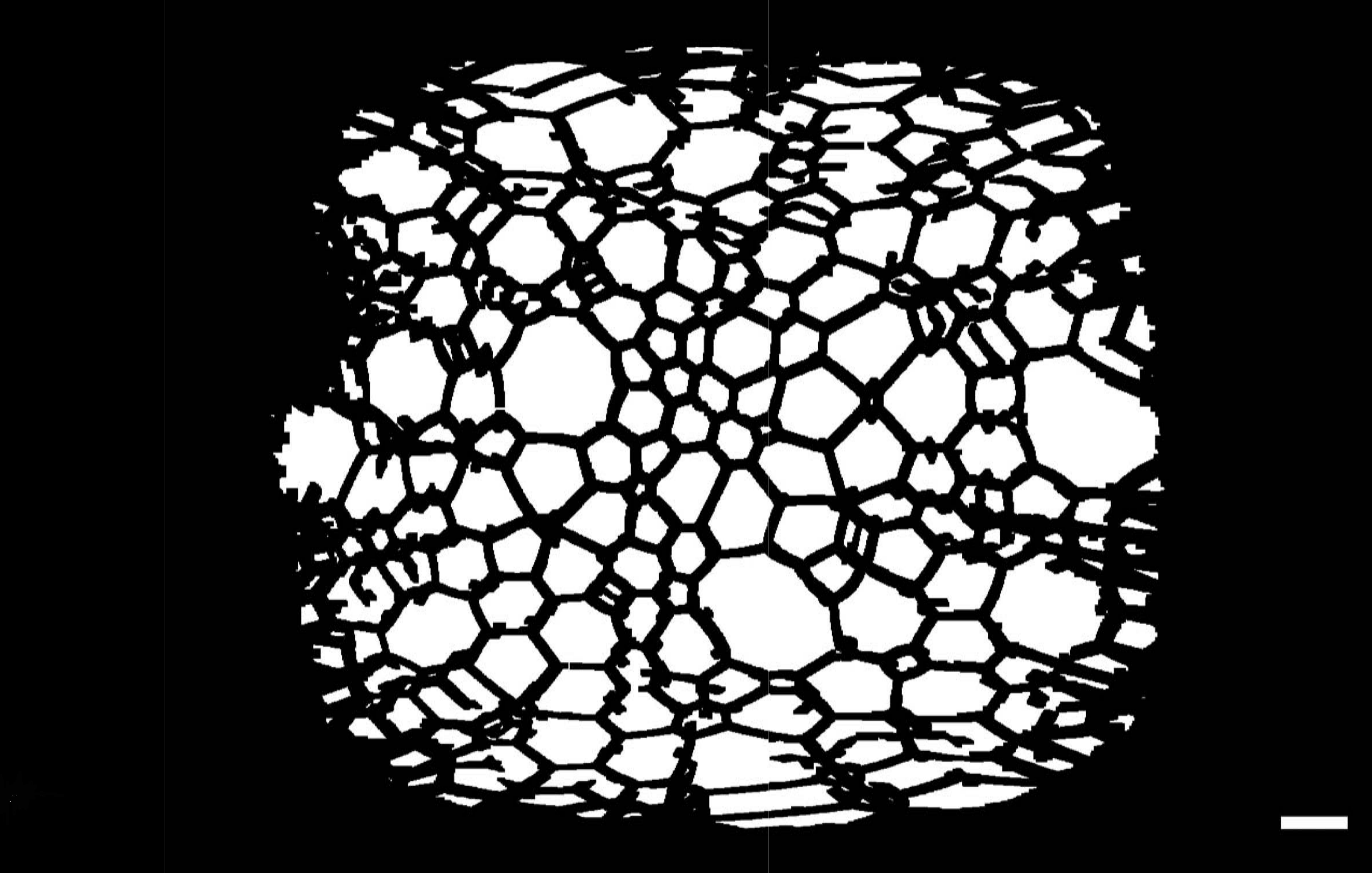}
\caption{Transformation of the photograph from Fig.~\ref{dome} by the Lambert equal area projection, to get the true areas of the cells.  There are errors in this result, but for a region in the center, typically about $4.5\times4.5~{\rm cm^{2}}$, it works well and the areas of these cells can be tracked.  This image is skeletonized before any actual areas are measured.  The scale bar is 1~cm.}
\label{newdome}
\end{figure}

After we have accounted for the fact that the Plateau borders on both the inner and outer domes are visible, we can then consider the areas of the individual bubbles.  The binary image with the correct latitudes and longitudes of the Plateau borders has errors, but does well for a region of interest in the center.  This resulting image, however, does not preserve the areas of the cells.  A simple projection that will preserve areas is the Lambert cylindrical equal area projection.  This projection is defined by \cite{MapBook}:
\begin{equation}
\begin{array}{l}
x = R \lambda\\
y = R \sin\varphi
\end{array}
\label{lambert}
\end{equation}
where as before $\lambda$ is the longitude and $\varphi$ is the latitude.  The $R$ used here is the average of the two domes used.  This produces an image with cells that have distorted shapes, but have the same areas as the actual cells on the curved surface.  This allows us to track individual bubble areas over time.  The result of this image analysis can be seen in Fig.~\ref{newdome}.

\section{Bubble Dynamics}

Using the method of finding the correct areas of individual bubbles described above, it is possible to track the areas of individual bubbles over time.  The method of identifying the correct Plateau borders and finding the correct areas sometimes has errors of failing to identify a Plateau border or adding an extra one, especially farther from the center, where distortion is greater.  It was possible to find bubbles, especially near the center, that would be correctly identified for a significant length of time.  The viewing region where bubbles can be tracked is typically about $4.5 \times 4.5~{\rm cm^{2}}$.  Not all bubbles in this region will necessarily be able to be tracked, but it is very rare for a bubble outside this region to be trackable.  The projections were compared to the original images, to ensure that the areas of the tracked bubbles were correct.  Images were taken at 200~minute intervals and analyzed to get the correct bubble areas.  Correctly identified bubbles were labeled and tracked over this period.  From this it was possible to get area versus time graphs for many bubbles.

\begin{figure}
\includegraphics[width = 3.00in]{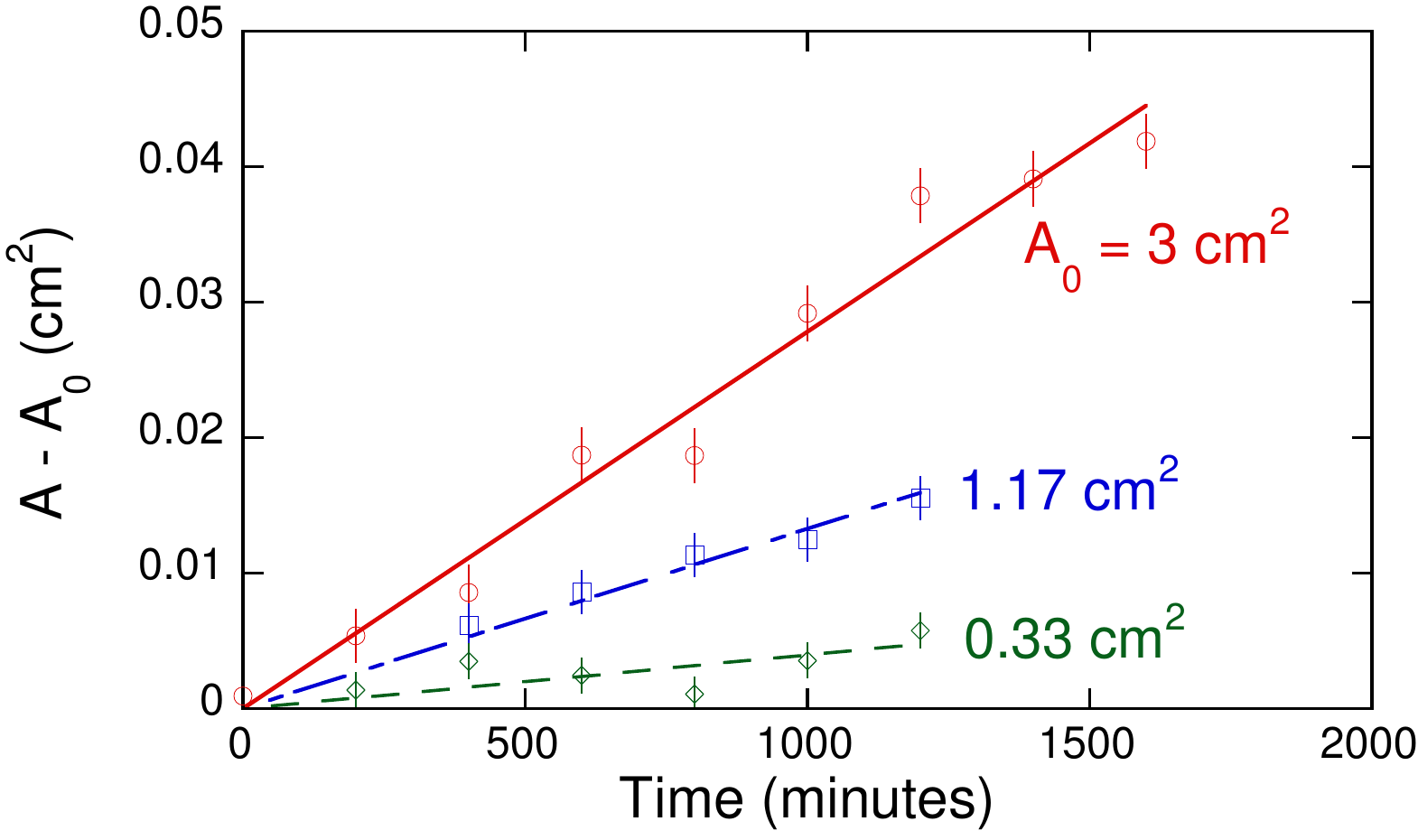}
\caption{(Color online) Area versus time for three different six-sided bubbles, with initial area, $A_0$, as labeled.  Lines are a linear fit to the data, from which $A_0$ was found and subtracted from the data.  The vertical error bars represent perimeter times pixel size divided by the square root of number of points in the perimeter; this statistical uncertainty matches the scatter in the data points.}
\label{growthvarea}
\end{figure}

We can now consider the form that the area versus time traces for the bubbles should take.  Avron and Levine's prediction for the coarsening of foam on a spherical substrate of radius $R$, Eq.~(\ref{AvronLevine2}), is a linear differential equation that can be solved analytically for area versus time:
\begin{eqnarray}
A &=& \left[ A_o + {\pi R^2 \over 3}\left(n-6\right)\left(1-e^{-{3K_ot \over \pi R^2}} \right) \right]  e^{3K_ot \over \pi R^2} \label{solution} \\
    &=& A_o+K_o t \left(n-6+{3A_o\over \pi R^2} \right)\left( 1 + {3K_o t \over 2\pi R^2} +\ldots  \right)   \label{expansion}
\end{eqnarray}
Here $A_o$ is the area of a bubble at time zero, and Eq.~(\ref{expansion}) is the Taylor expansion in $(K_ot/R^2)$.  Note that the standard von~Neumann result, $A=A_o+K_o(n-6)t$, is obtained in the limit $R\rightarrow\infty$.

Examples data for the area versus time of three six-sided bubbles with different initial areas are plotted in Fig.~\ref{growthvarea}.  There the initial area of each bubble was subtracted off so that the traces are easily comparable.  The lines are a linear fit to the data, giving a constant growth rate ${\rm d}A/{\rm d}t$ that is positive.  It is possible to fit the data to the full exponential form of Eq.~(\ref{solution}), but the additional terms in the expansion from Eq.~(\ref{expansion}) are much, much smaller than the linear term, so it is sufficient to fit to an ordinary line.  The key features of Fig.~\ref{growthvarea} is that the six-sided bubbles grow, and that the larger ones grow faster.  This agrees with Avron and Levine, and contrasts strongly with the case of a flat sided cell, where six-sided bubbles neither grow nor shrink according to the usual von~Neumann equation.  In particular, in our recent experiments on the coarsening of bubbles in a flat cell \cite{RothDurian11}, where the liquid fraction was varied, the six-sided bubbles all had growth rates scattered around zero to within statistical uncertainty.

\begin{figure}
\includegraphics[width = 3.00in]{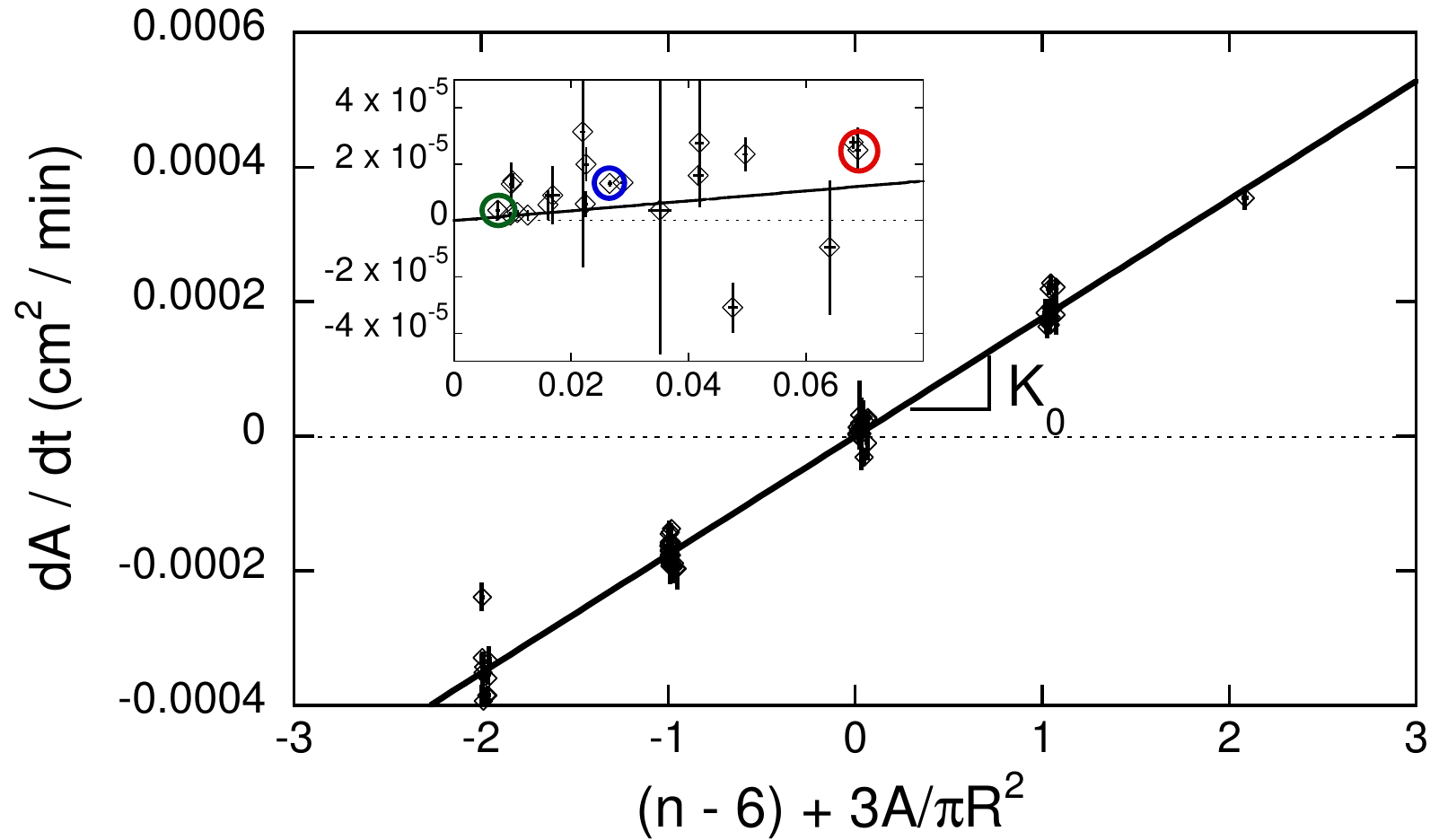}
\caption{(Color online) Growth rate of bubble area versus the expected factor from Avron and Levine, Eq.~(\ref{AvronLevine2}), where $A$ is bubble area and $R$ is the radius of the dome.  The error bars represent the uncertainty in slopes of linear fits to bubble area versus time.  The solid line is a proportionality fit to all data,  with slope $K_o = 0.00018\pm0.00008~{\rm cm^{2} / min}$.  The inset shows a blow-up of the data for the $n=6$ sided bubbles.  There, the growth rates for the three representative bubbles of Fig.~\ref{growthvarea} are circled in color.}
\label{domevn}
\end{figure}

We now measure the growth rate for {\it all} bubbles, as illustrated in Fig.~\ref{growthvarea}, and we compare to the expected relationship from Avron and Levine's modification to von~Neumann's law.  This is plotted in Fig.~\ref{domevn}, where the y-axis is the coarsening rate, and the x-axis is the expected proportionality for a dome of constant curvature given by Eq.~(\ref{AvronLevine2}).  Each point represents a single bubble.  The line is a proportionality, with slope $K_o$, which is the only fitting parameter.  The inset is a blow-up showing all the six-sided bubbles, and highlighting the three bubbles featured in Fig.~\ref{growthvarea}.  For six-sided bubbles the growth rates are all positive, except for one or two outliers.  There is notable scatter, but the evident trend is that ${\rm d}A/{\rm d}t$ increases with bubble size. 

\begin{figure}
\includegraphics[width = 3.00in]{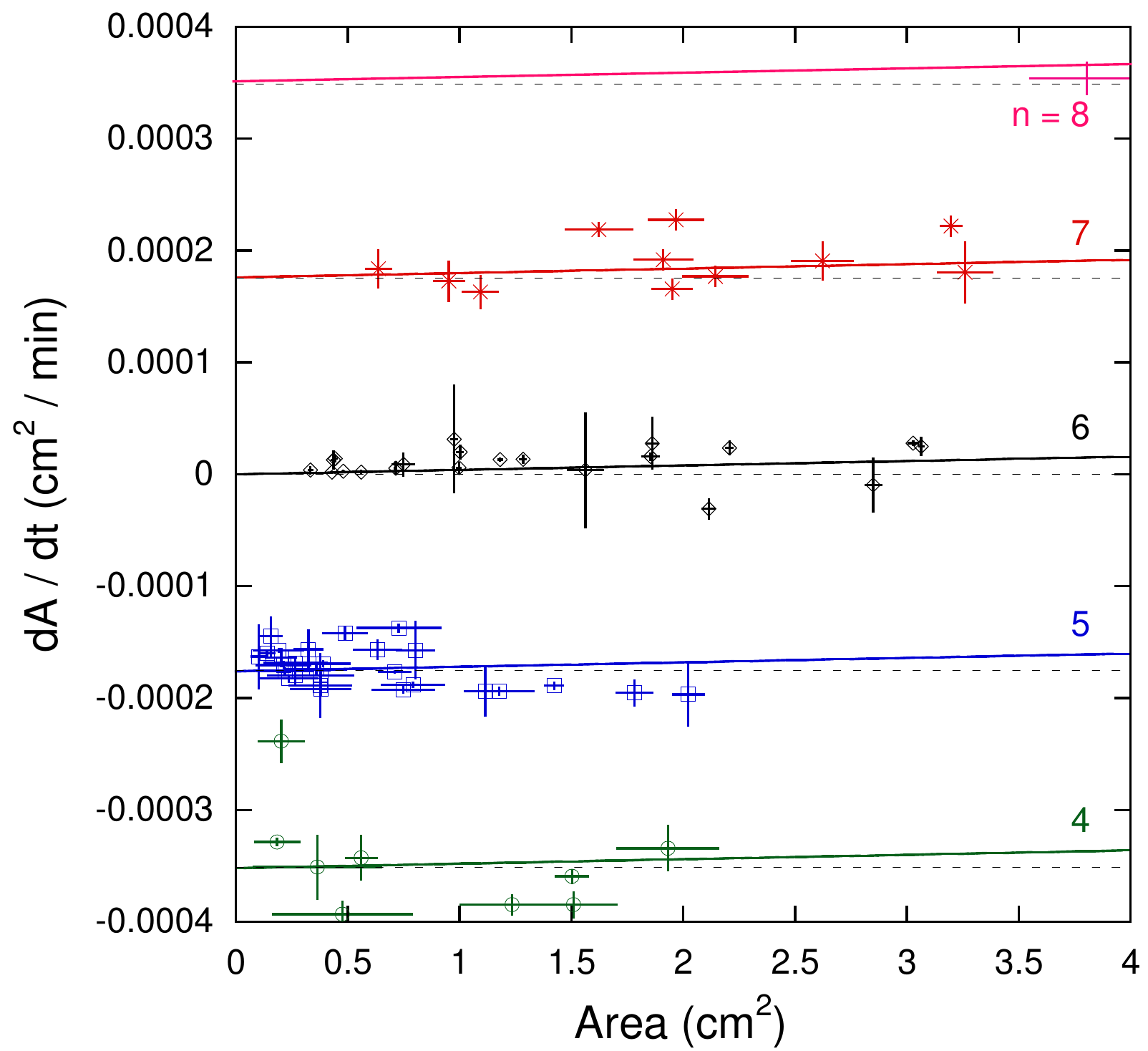}
\caption{(Color online) Growth rate of bubbles versus area.  Symbol types distinguish bubbles with different number $n$ of sides, as labeled.  Horizontal lines are the expected relationship from the uncorrected von~Neumann law, shown in Eq.~(\ref{VN}), using $K_o$ from Fig.~\ref{domevn}.  Solid lines are the expected relationship using the corrected von~Neumann law from Avron and Levine, shown in Eq.~(\ref{AvronLevine2}), using $K_o$ from Fig.~\ref{domevn}.  The data and vertical error bars are the same as shown in Fig.~\ref{domevn}.  The horizontal error bars represent the range of area over which each bubble grew with the specified rate and number of sides.}
\label{domecoarsening}
\end{figure}

Another way to compare growth rate data to Avron and Levine is to plot the coarsening rate against area, as shown in Fig.~\ref{domecoarsening}.  Here each point represents a single bubble, color coded by the number of sides.  The horizontal lines are the expected relationship from the unmodified von~Neumann's law, as seen in Eq.~(\ref{VN}), using the same constant of proportionality $K_o$ as found in Fig.~\ref{domevn}.  The solid lines are the expected relationship from the modified von~Neumann's law, as seen in Eq.~(\ref{AvronLevine2}), again using the same value of $K_o$.  We see that the data are generally consistent with the predicted modification.  This is most evident in the six-sided bubbles, which are virtually all growing.  The rate of area change also appears to increase with area for $n = 7$, but is masked by noise for other $n$.

\section{Bubble Distributions}

With our system it was also possible to measure distributions such as $p(n)$, the probability that a bubble has $n$ sides, and $m(n)$, the average number of sides of the neighbors of an $n$-sided bubble.  Unlike the case of the flat cell, we do not expect to reach a scaling state where these statistical quantities remain constant over time.  Because the growth rate of a bubble grows with its area, we expect at long times to have large bubbles grow rapidly to dominate the system, and this will cause bubble statistics and distributions to change with time.  Our system is at much earlier times, where the modification to a bubble's growth rate due to its area is small.  This modification still should have some impact and we do indeed find that our statistics deviate from the ordinary scaling state observed in the flat cell.

\begin{figure}
\includegraphics[width = 3.00in]{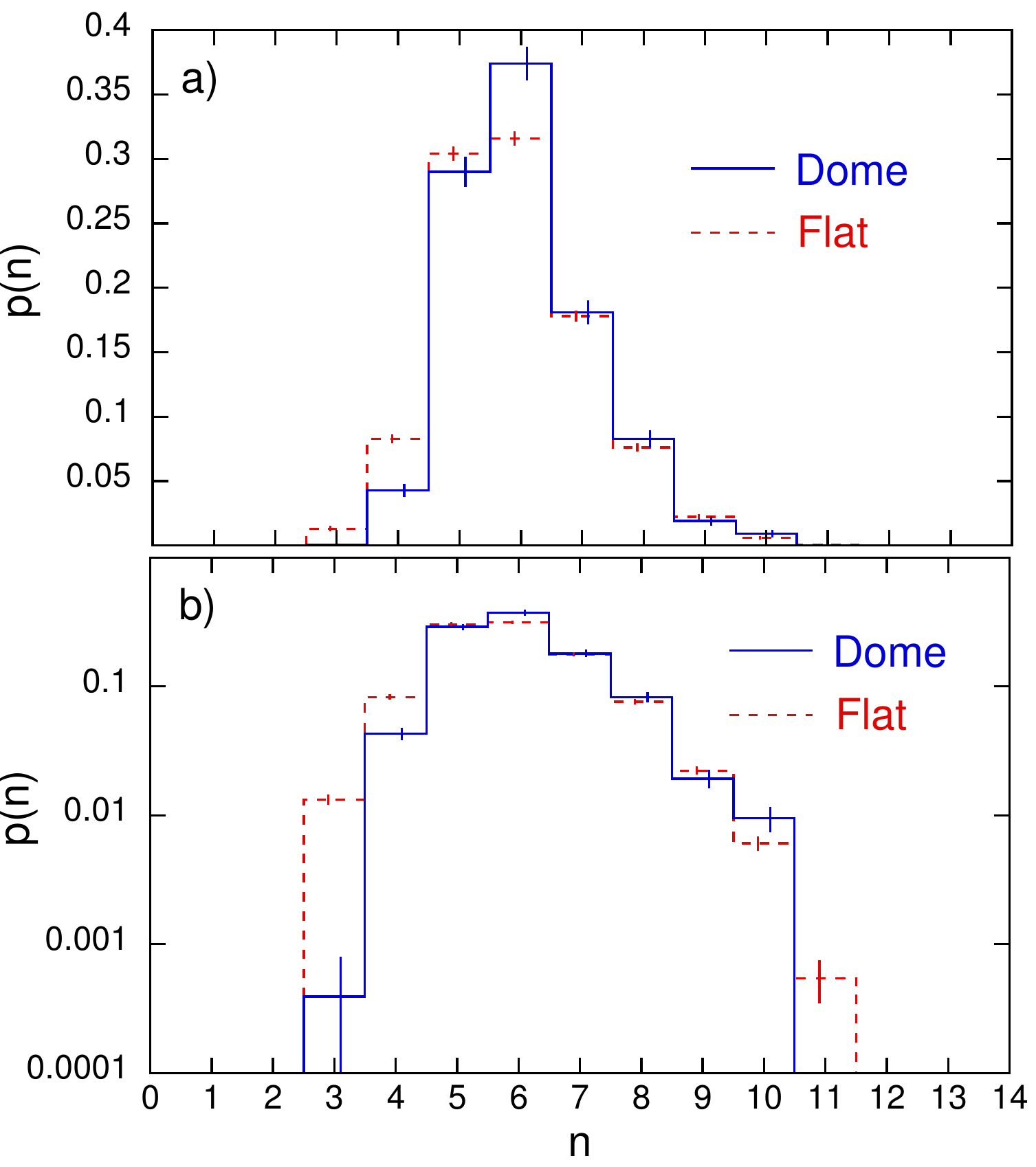}
\caption{(Color online) Side number distribution for a flat cell and for the dome.  For the flat cell, the average number of sides is $\langle n \rangle = 5.92 \pm 0.01$ and the variance $\mu_{2} = 1.56 \pm 0.02$.  For the dome $\langle n \rangle = 6.06 \pm 0.1$ and $\mu_{2} = 1.30 \pm 0.05$. Data for flat cell is taken from Ref.~\cite{RothDurian11}.  The vertical error bars are from a fractional area of one over the square-root of number of bubbles with specified side number.}
\label{domeandflat}
\end{figure}

To measure our statistics, a $8~\textrm{cm} \times 8~\textrm{cm}$ region of interest was defined in the center of the dome, and the number of sides of all bubbles completely or partially within this region was recorded by hand.  This was done for photographs at 400~minute intervals from the earliest photograph of a two dimensional foam to the first occurrence of a rupture, for a total period of typically 48~hours.  This was repeated for three runs.  This data was used to produce a distribution of the number of sides, which can be seen in Fig.~\ref{domeandflat}.  Also shown for comparison is the distribution found for a flat two dimensional foam in Ref.~\cite{RothDurian11}.  We see that as compared to the flat cell, the dome has a surplus of six-sided bubbles, and a deficit of 3, 4, and 5 sided bubbles.

It is also possible to describe these distributions by their average, $\langle n \rangle = 6.06 \pm 0.1$, and by their variance, $\mu_{2} = \sum p(n) (n-6)^{2}$.  We measure the variance to be $\mu_{2} = 1.30 \pm 0.05$.  This value is lower than was measured for the flat cell, $\mu_{2} = 1.56 \pm 0.02$ \cite{RothDurian11}, indicating that the width of the distribution is narrower.

\begin{figure}
\includegraphics[width = 3.00in]{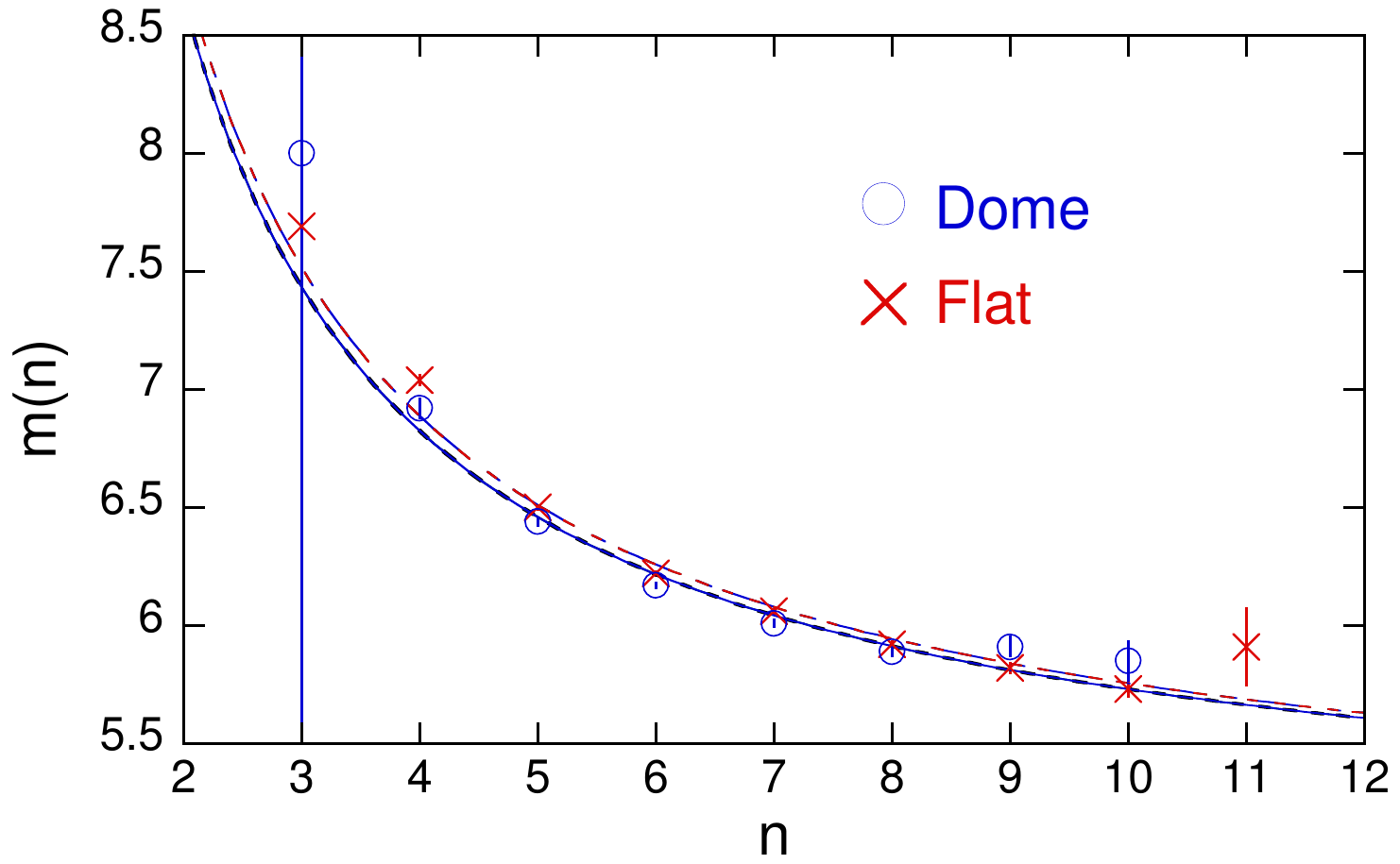}
\caption{(Color online) Aboav-Weaire relationship for a flat cell and for a dome.  The y-axis is $m(n)$, the average number of sides of an $n$ sided bubble.  Lines are fits to the Aboav-Weaire law, $m(n) = 6-a + [(6a - \mu_{2})/n]$, where $a$ is the only fitting parameter.  We find $a = 1.1 \pm 0.08$ for the flat cell and $a = 0.96 \pm 0.09$ for the dome \cite{RothDurian11}.  The variances, $\mu_{2}$, for these systems can be found in the caption to Fig.~\ref{domeandflat}.}
\label{mn}
\end{figure}

From this same data we can measure $m(n)$, the average number of sides of the neighbors of an $n$ sided bubble.  We expect $m(n)$ to be related to $n$ by the Aboav-Weaire law, which predicts $m(n) = 6-a + [(6a - \mu_{2})/n]$.  In this equation $a$ is a fitting parameter which is usually found to be around 1.  Our measurements for $m(n)$ can be seen in Fig.~\ref{mn} along with the measurements of $m(n)$ for a flat cell.  Fits to the Aboav-Weaire law are also shown, using the relevant value for $\mu_{2}$ in each case.  We find $a = 1.1 \pm 0.08$ for the flat cell \cite{RothDurian11} and $a = 0.96 \pm 0.09$ for the dome.  We see that in both cases there appears to be more curvature in the data than predicted.  The data for the flat cell also seems to fit the form better than for the dome.

\section{Conclusion}

In this experiment we were able to measure both bubble statistics and bubble dynamics of a foam on a curved two-dimensional surface of radius $R$.  For bubble statistics we find that bubbles with few sides are under-represented as compared to a two dimensional foam in flat space.  We also find that the Aboav-Weaire law generally holds, though not quite as well for the dome as for the flat cell.  For bubble growth rates, in general, it is difficult to observe the effect of the term added to von~Neumann's law by Avron and Levine to account for substrate curvature.  Our maximum bubble size is around $A_{max} = 3.5~{\rm cm^{2}}$, as compared to $R^{2} = 41.6~{\rm cm^{2}}$; therefore, for all our bubbles $3A / (\pi R^{2}) \ll |n-6|$ holds, except for $n = 6$.  This is why all the data in Fig.~\ref{domevn} lie at $x$ values near $(n-6)$.   Even if we managed to get a single bubble of ${\rm 20~cm^{2}}$ to completely fill our viewing area, the rate of area change would be ${\rm d}A / {\rm d}t = K_o [(n - 6) + 0.46]$, so that a seven sided bubble would have less than a $50\%$ increase in growth rate.  For these reasons, the clearest signal we see of curvature effects is that six-sided bubbles systematically grow and do so faster for larger bubbles.  The coarsening data as a whole is consistent with Avron and Levine's modification to von~Neumann's law to account for coarsening on a curved surface.

%=========================================================================================

\begin{acknowledgments}
We thank J.~Rieser for work on preliminary stages of this experiment.  This work was supported by NASA Microgravity Fluid Physics Grant NNX07AP20G
\end{acknowledgments}

\bibliography{CoarseningRefs}

\end{document}